\documentclass{ws-procs10x7}

\graphicspath{{ps/}}

\begin{document}

\title{Particle Detector R\&D}

\author{M.V.Danilov}

\address{Institute for Theoretical and Experimental Physics,
  B.Cheremushkinskaya 25, 117218 Moscow,
RUSSIA\\E-mail: danilov@itep.ru}

\twocolumn[\maketitle\abstract{
Recent results on the particle detector R\&D for new accelerators are reviewed.
Different approaches for the muon systems, hadronic and electromagnetic calorimeters,
particle identification devices, and central trackers are discussed.
Main emphasis is made on the detectors for the International Linear Collider and 
 Super B-factory. A detailed description of a novel photodetector, a so called 
Silicon Photomultiplier, and its applications in scintillator detectors is presented. }]

\section{Introduction}

Particle detector R\&D is a very active field. Impressive results of
the long term R\&D for the Large Hadron Collider (LHC) are being
summarized now by the four LHC detector collaborations. A worldwide
effort is shifted to the detector development for the future
International Linear Collider (ILC) and for the Super B-factory. The
detector development for the FAIR facility has already
started. Several groups perform R\&D studies on detectors for the next
generation of the hadron colliders. 

This review is devoted mainly to the detector development for the ILC
and Super B-factory. The vertex detectors are not discussed here in
order to provide more details in other fields. R\&D on the vertex
detectors is very active and deserves a separate review. This review
is organized following the radius of a typical collider detector from
outside to inside.  

\section{Muon Detectors}

Muon detectors cover very large areas. 
Therefore they should be robust and inexpensive.
Resistive Plate Chambers (RPC) are often used in the present detectors,
 for example at the B-factories. 
 In the streamer mode RPCs provide large signals. 
Hence it is possible to use very simple electronics. 
Another advantage is a possibility to have different shapes of read
out electrodes that match best the physics requirements. 
For example the BELLE RPCs have ring and sector shaped readout electrodes in the end cap regions.

The European CaPiRe Collaboration developed a reliable industrial
 technique for the glass RPC production\cite{capire}.
 The production rate of more than 1000 square meters per day is possible. 
The RPC efficiency is larger than 95\% up to the counting rates of
$1Hz/cm^2$. 
 This is reasonably adequate for the ILC
 detector but at the Super-B factory one expects by far larger
 rates. The RPCs in the proportional mode can stand about hundred
 times higher counting rates.  

Scintillator strip detectors can work
 at even higher rates. A very attractive possibility is to use
 scintillator strips with Wave Length Shifting (WLS) fibers 
read out by so called Silicon Photo Multipliers (SiPM).

\begin{figure}[thb]
\centering
\begin{picture}(500,200)
\put(0,0){\includegraphics[width=7.2cm]{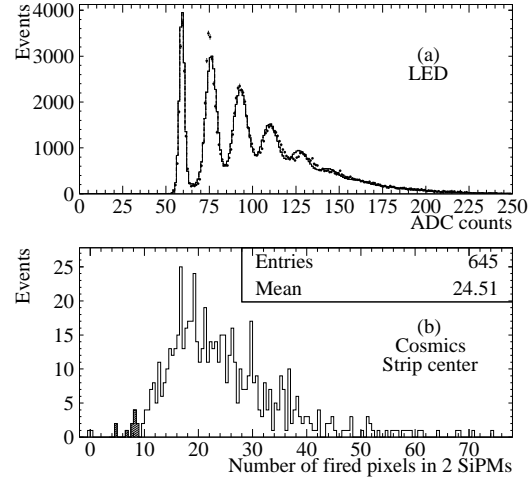}}
\end{picture}
\caption{(a) SiPM response to short weak LED flashes.
 The fit curve is a simple model of the SiPM response.
 (b) Number of fired pixels  in two SiPMs by a cosmic particle at the strip center.
 Few entries around zero belong to the pedestal.}
\label{strip}
\end{figure}

SiPM is a novel photo detector developed in Russia\cite{SiPM,MRSAPD,Sad}. 
It will be mentioned many times in this review.
 Therefore we shall discuss its properties in 
 detail~\footnote{Three groups developed such devices and produce them.
They use different names for their products. We will use a generic name SiPM
for all types of multipixel Si diodes working in the Geiger mode.
New types of SiPMs are being developed by several groups including Hamamatsu}.
 SiPM is a matrix of 1024 = 32$\times$32 independent silicon
 photodiodes~\footnote{SiPMs can be produced with different number of pixels
 in the range 500-5000. We describe here the SiPM used for the hadronic calorimeter prototype
for the ILC\cite{minical}} 
covering the area of $1\times 1\,mm^2$.
Each diode has its own quenching polysilicon resistor 
of the order of a few hundred k$\Omega$. 
All diode-resistor pairs, called pixels 
later on, are connected in parallel.
A common reverse bias voltage $V_{bias}$ is applied across them.
Its magnitude of the order of $40-60 V$ is high enough 
to start the Geiger discharge if any free charge carrier appears in the $p-n$ junction 
depletion region. 
The diode discharge current 
causes a voltage drop across the resistor.
 This reduces the voltage across the diode below the breakdown voltage $V_{breakdown}$ and
the avalanche dies out. One diode signal is 
$Q_{pixel}=C_{pixel}(V_{bias}-V_{breakdown})$ where $C_{pixel}$ is the
 pixel capacitance. 
Typically $C_{pixel}\sim 50$~fF and
$\Delta V = V_{bias}-V_{breakdown}\sim 3 V$ yielding $Q_{pixel}\sim 10^6$ electrons. 
Such an amplification is similar to the one of 
a typical photomultiplier and 3--4 orders of magnitude larger than the 
amplification of an Avalanche Photo Diode (APD) working in the proportional mode. 
$Q_{pixel}$ does not depend on the number of primary carriers
 which start the Geiger discharge. 
Thus each diode detects the carriers created e.g. by a photon, 
a charged particle or by a thermal noise with the same response 
signal of $\sim 10^6$ electrons. 
Moreover the characteristics of different diodes inside the SiPM are also very similar. 
When fired, they produce 
approximately the same signals. 
This is illustrated in Fig.~\ref{strip}a. 
It shows the SiPM response spectrum when it is illuminated by weak flashes of 
a Light Emitting Diode (LED). 
First peak in this figure is the pedestal. 
The second one is the SiPM response when it detects exactly one photon. 
It is not known which diode inside the SiPM
produces the signal
since all of them are connected to the same output. 
However since the responses of all pixels are similar, 
the peak width is small.
If several pixels in the SiPM are fired, the net charge signal
 is the sum of all charges.
The third, forth and so on peaks in Fig.~\ref{strip}a 
correspond to 2, 3, ... fired pixels.

The SiPM photodetection efficiency depends on the light wave length 
and the overvoltage $\Delta V$.
A typical value is about 10-15\% for the green light. 
It includes geometrical inefficiency due to dead regions
 in the SiPM between the pixels. 
Thus SiPM and traditional photomultipliers have similar gain and efficiency.
However, SiPM is approximately twice cheaper than one channel
 in the multianode photomultiplier and further cost reductions are expected in 
case of mass production.
SiPM can work in the magnetic field, 
so there is no need in the light transportation  
out of the magnetic field.
SiPM is so tiny that it can be mounted directly on the detector. 
This minimizes the light losses because of a shorter fiber length. 
SiPM has a quite high noise rate of about 2~MHz at 0.1 photoelectron threshold. 
However the noise rate drops fast with increasing threshold. 

Fig.~\ref{strip}b shows the pulse height spectrum for cosmic particles obtained 
with the scintillator strip detector read out by two SiPMs~\cite{balagura}.
The detector consists of a $200\times2.5\times1\,cm^3$ plastic scintillator strip 
and a wavelength shifting fiber 
read out by two SiPMs installed at the strip ends.  
 The strip is extruded 
from the granulated polystyrene with two dyes
 at the ``Uniplast'' enterprise in Vladimir, Russia. 
The Kuraray multicladding WLS fiber Y11 (200) with 1~mm diameter
 is put in the 2.5~mm deep groove in the middle of the strip.
No gluing is used to attach the WLS fiber to the SiPM or to the strip. 
There is about 200~$\mu$m air gap between 
the fiber end and the SiPM.
To improve the light collection efficiency, the strip is wrapped 
in the Superradiant VN2000 foil produced by the 3M company. 
 
In the worst case when the particle passes through the strip center
there are 13.7 detected photons. 
The Minimum Ionizing Particle (MIP) signal in Fig.~\ref{strip}b 
is well separated from the pedestal. 
The detector efficiency averaged over the strip length is 
as high as $99.3\pm 0.3\%$ at the 8 pixel threshold. 
Such a threshold is sufficient to reduce the SiPM noise rate to $5kHz$.

The ITEP group has also studied $100\times4\times1\,cm^3$ strips~\cite{balagura-privat}
 with a SiPM~\cite{MRSAPD} 
at one end of the WLS fiber 
and a 3M Superradiant foil mirror at the other end.
 The strips were produced by the extrusion technique in Kharkov.
 The strip surface was covered by a Ti oxide reflector co-extruded 
together with the strip.  The Kuraray Y11, 1mm diameter fiber was glued into
 the 3mm deep groove with an optical glue. 
SiPM was also glued to the fiber. 
More than 13 photoelectrons per MIP  were detected
 at the strip end far from the SiPM. 
With such a large number of photoelectrons the efficiency 
of more than 99\% for MIP was obtained with the threshold 
of 7 photoelectrons. The detector can work at counting rates above $1kHz/cm^2$.
 This is sufficient for the Super B-factory. 
Therefore the Belle Collaboration plans to use this technique for
 the $K_L$ and muon system upgrade in the end cap region~\cite{MVDKEK}.  

A scintillator tile structure can be used for even higher rates. 
Sixteen $10\times10\times1 \,cm^3$ tiles read out by two SiPM~\cite{MRSAPD} each were
 tested at the KEK B-factory~\cite{MVDKEK}.
 They demonstrated a stable performance adequate for the Super B-factory. 
An eight square meter cosmic test system for ALICE TOF RPC chambers
 is constructed at ITEP~\cite{alicetest}.
 It consists of $15\times15\times1\,cm^3$ tiles read out by two SiPMs~\cite{MRSAPD} each.
 The counters have an intrinsic noise rate below $0.01\,Hz$, 
the time resolution of $1.2\,nsec$, and the rate capabilities up to $10\,kHz/cm^2$.

\section{Hadronic Calorimeters}

 The precision physics program at the future International Linear Collider (ILC) 
 requires to
 reconstruct heavy bosons (W,Z,H) in hadronic final states in multijet
 events. In order to do this a jet energy resolution of better than
 30\%/$\sqrt{E}$ is required~\cite{felix}.
 The energy E is measured in GeV in this expression and in similar expressions 
for the energy resolution below. 
 Monte Carlo (MC)
 simulations demonstrate that such a resolution can be achieved using
 a novel "particle flow" (PF) approach in which each particle in a jet
 is measured individually~\cite{Morgunov}. Momenta of charged
 particles are determined using tracker information. Photons are
 measured in the electromagnetic calorimeter (ECAL). Only neutrons and $K_L$
 should be measured in the Hadronic calorimeter (HCAL). They carry on
 average only about 12\% of the jet energy. Therefore the HCAL can have 
 modest energy resolution. The major problem is to reconstruct showers
 produced by charged tracks and to remove the corresponding energy
 from the calorimetric measurements. This requirement makes the pattern
 recognition ability to be a major optimization parameter of HCAL.

The CALICE Collaboration investigates two approaches for the HCAL. 
In the digital approach only one bit yes/no information is recorded for each cell.
Extremely high granularity of about $1 \,cm^2$/cell is required in this case.
In the analog approach the pulse height information is recorded for each cell.
However a very high granularity of about $5\times5 \,cm^2$/cell is still required~\cite{MVDParis}.
Such a granularity practically can not be
 achieved with a conventional readout approach with WLS fiber 
and a multianode photomultiplier (MAPM). 
The use of tiny SiPMs makes such a granularity achievable.

\subsection{Analog Hadronic Calorimeters}

A small 108 channel hadronic calorimeter prototype has been built  in order to
gain experience with this novel technique~\cite{minical}. 
The calorimeter active modules
have been made at ITEP and MEPhI. Scintillator
tiles are made of a cheap Russian scintillator using a molding
technique. A Kuraray Y11 1mm diameter double clad WLS fiber is
inserted into a 2mm deep circular groove without gluing. The SiPM is
placed directly on the tile and occupies less than 0.5\% of a
sensitive area. 
There is an air gap of about 100$\mu m$ between the fiber and SiPM. 
Signals from SiPMs are sent directly to 
LeCroy 2249A ADCs via 25 meter long 50~$\Omega$ cables.

\begin{figure}[ht]
\centering
\begin{picture}(500,140)
\put(20,-35){\includegraphics[width=7.5cm]{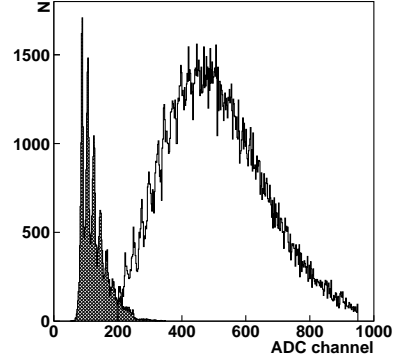}}
\end{picture}
\caption{Pulse height spectrum from a tile with SiPM for low intensity LED light
 (hatched histogram)
 and for MIP signals from a $\beta$-source.}
\label{spectrum}
\end{figure}

A lot of
R\&D has been performed in order to increase the light yield and the
uniformity of the response. For better light collection
the surface of the tiles is covered with 3M Superradiant foil. The
tile edges are chemically treated in order to provide diffuse light
reflection and separation between tiles. A light yield of more than
20 photoelectrons per MIP has been achieved for $5\times5\times 0.5\,cm^3$
tiles. Fig.~\ref{spectrum} shows LED and \mbox{$\beta$-source} ($^{90}$Sr)
signals from such a tile. Peaks with different number of
photoelectrons are clearly seen. Signals from the
$\beta$-source are very similar to MIP signals. 

\begin{figure}[ht]
\centering
\begin{picture}(550,140)
\put(5,-3){\includegraphics[width=6.3cm]{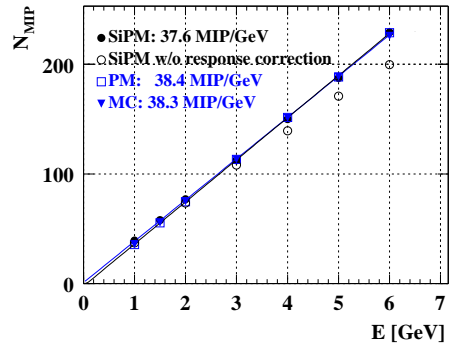}}
\end{picture}
\caption {Calorimeter response normalized to number of MIPs versus
beam energy; solid points (open circles) show SiPM data with (without)
response function correction, squares are MAPM data and triangles are
MC predictions . }
\label{linearity}
\end{figure}

The HCAL prototype was successfully operated at the DESY electron test
beam.  Fig.~\ref{linearity}
shows the linearity of the calorimeter response measured with
SiPM (circles) and 
MAPM (squares). The agreement between two
measurements is better than 2\%. The linear behavior of the SiPM
result (better than 2\%) demonstrates that the applied saturation correction 
due to limited number of pixels in the SiPM 
 is reliable. 
The obtained energy resolution agrees well with MC expectations and with a
resolution obtained using conventional MAPMs as well as APDs~\cite{APD}.
 The obtained resolution of about $21\%/\sqrt{E}$ is modest since this
is a hadron calorimeter. 

The 8000 channel HCAL prototype with the SiPM readout is being constructed
 by a subgroup of the CALICE Collaboration~\cite{sefkow05}. 
The $3\times 3 \,cm^2$ tiles are used in the central part of the calorimeter 
in order to test a semidigital approach. 
MC studies predict that the calorimeter with $3\times 3 \,cm^2$ cells 
and the 3 threshold measurement of the energy deposited in the tile 
should provide as good performance as a 
digital (yes/no) calorimeter with the $1\times 1 \,cm^2$ granularity~\cite{Zutshi}.

\subsection {Digital Hadronic Calorimeters}

\begin{figure}[ht]
\centering
\begin{picture}(550,165)
\put(5,-3){\includegraphics[width=6.2cm]{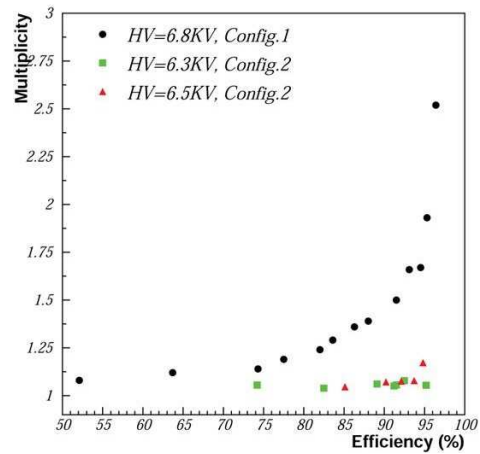}}
\end{picture}
\caption {Pad multiplicity dependence on the efficiency for two types of RPC:
 full circles - standard RPC with two glass sheets;
 triangles and squares - the RPC with one glass sheet. }
\label{USRPC}
\end{figure}

The RPC based digital HCAL is developed by a subgroup
 of the CALICE collaboration~\cite{repond}. 
They studied several RPC geometries and gases in order
 to optimize the efficiency and to reduce the cross-talk between pads. 
Fig.~\ref{USRPC} shows the efficiency and pad multiplicity due to 
cross-talk obtained with the developed RPC prototype.
 The prototype consists of two sheets of  floating glass 
with the resistive paint layer (1Mohm/square)
 and the gas gap of 1.2 mm. In works in the proportional mode 
and has the efficiency above $90\%$ up to the rates of  $50\,Hz/cm^2$.
 The pad multiplicity is about~1.5.
 Much smaller pad multiplicity is observed in the RPC
 in which the readout electrode defines the gas sensitive volume 
instead of the glass sheet (see Fig.~\ref{USRPC}). 
It will be interesting to study further the properties of this promising RPC.

The GEM based digital HCAL is studied by another subgroup
 of the CALICE Collaboration~\cite{UTA}.
 They developed a procedure for the large area double 
GEM chamber production. 
A small prototype demonstrates the 95\% efficiency at 40mV threshold 
and the pad multiplicity of 1.27.   
The 3M company plans to produce already in 2005 very long GEM foils of about $30\,cm$ width.

The number of channels in the digital HCAL is enormous.   
Therefore cheap and reliable electronics is the key issue for this approach.
 The RPC and GEM digital HCAL teams
 develop jointly the electronics suitable for both techniques.

\subsection {The DREAM Calorimeter}

Usually calorimeters have different response to electromagnetic 
and hadronic showers of the same energy. 
Therefore the fluctuations of the electromagnetic energy fraction
 in the hadron shower is one of the main reasons
 for the deterioration of the energy resolution.

 In the Dual Readout Module (DREAM) calorimeter~\cite{wigmans}
 the electromagnetic energy in the hadronic shower is measured
 independently using quartz fibers sensitive
 only to the Cherenkov light produced dominantly by electrons. 
The visible energy is measured by scintillation fibers. 
The electromagnetic energy fraction in the shower can be determined 
by the comparison of the two measurements. 
This allows to correct for the different calorimeter response 
to the electromagnetic showers and to improve the energy resolution.
 A very similar response to electrons, hadrons, and jets was obtained in the 
DREAM calorimeter prototype after this correction. 
The ultimate energy resolution of the DREAM calorimeter
 is expected to be better than $30\%/\sqrt{E}$. 
Unfortunately the shower leakage and insufficient amount of the Cherenkov light
 limited the measured prototype calorimeter resolution to $64\%/\sqrt{E}$ only.

The fluctuations of the visible energy because of the nuclear energy loss
 can be corrected for by adding to the DREAM structure the third type of fibers
 sensitive to neutrons. 
In this case the ultimate energy resolution of $15\%/\sqrt{E}$ is expected~\cite{wigmans}.
 There are many  nice ideas how to separate different mechanisms
 in the hadronic shower and to improve the energy resolution~\cite{wigmans}. 
However they should be first demonstrated experimentally.
\section{Electromagnetic Calorimeters}
\subsection{Electromagnetic Calorimeters for ILC}

The requirement of a high granularity for the ILC detectors 
leads to the choice of very dense electromagnetic calorimeters 
with a small Mollier radius ($R_M$). 
Silicon/tungsten, scintillator/tungsten and scintillator/lead 
sandwich options are developed. 
The price for the high granularity is a 
modest energy resolution of the proposed calorimeters.
 
The CALICE collaboration constructs the Si/W prototype with about 10 thousand
 channels~\cite{emcalice}.  
The pad size is as small as $1\times1\,cm^2$. The Si thickness is $500\mu m$.
The tungsten plate thickness is  $1.4\,mm$, $2.8\,mm$, and $4.2\,mm$ 
in the front, middle, and rare parts of the calorimeter. 
One third of the prototype has already been tested at the DESY electron beam 
and demonstrated a stable behavior. 
The signal to noise ratio of 8.5 was obtained for  MIP. 
The tests of the whole calorimeter will start this Winter. 
The combined tests with the analog hadronic calorimeter are planned in 2006 as well.

The detector and readout plane thickness is $3.4\,mm$ in the present prototype. 
It will be reduced to $1.75\,mm$ including the readout chip 
in the final design resulting in $R_M=1.4\,cm$. 

The US groups (SLAC, UO, BNL) develop even more aggressive design 
of the Si/W calorimeter~\cite{usecal} for the small radius Si based ILC detector (ILC SiD). 
The detector and readout plane thickness is $1\,mm$ only
which results in the $R_M=1.4\,cm$. 
Together with HPK they developed the Si detector consisting of 1024 
hexagonal pads with $5\,mm$ inner diameter.   
The detector is read out by a specially developed electronic chip~\cite{marti}.
The measured MIP signal in this detector is 26k electrons while the pedestal 
width is 780 electrons. 
The Si/W calorimeter for ILC is also developed in Korea~\cite{koreaecal}.

A hybrid scintillator/lead calorimeter prototype with three Si layers 
has been built and tested by the INFN groups~\cite{itecal}. 
The $5\times5\times0.3\,cm^3$ scintillator tiles are combined into 4 longitudinal sections.
Three layers of $9\times9mm^2$ Si pads are placed between the sections at 2, 6, and 12$X_0$.
 The prototype demonstrated a good energy resolution of $11.1\%/\sqrt{E}$.
 It has the impressive spatial resolution of $2mm$ at $30GeV$ and $e/\pi$ 
rejection below $10^{-3}$. 
However it is not clear whether the granularity is sufficient for the PF method. 
Also the light transportation in the real detector will be extremely difficult. 
The use of SiPMs can solve the last problem.

The Japan-Korea-Russia Collaboration
 develops a scintillator/lead  calorimeter 
with the SiPM readout~\cite{japecal}. 
The active layer consists of two orthogonal planes of 
$200\times 10\times 2mm^3$ scintillator strips and a plane of $40\times40\times2 mm^3$
 tiles with WLS fibers.
 The fibers are readout by 
SiPMs developed at Dubna~\cite{Sad}.  Even shorter strips of $40\times 10\times2mm^3$
 are considered as an alternative. 
The signal of 5p.e./MIP was obtained with the $200\times 10\times 2mm^3$ strips.

\subsection{Electromagnetic Calorimeters for the Super B-Factory}

Electromagnetic calorimeters for the Super B-factory should have a very good energy resolution
 and a fast response. 
They should be  radiation hard up to about 10 kRad in the endcap region.
The present CsI(Tl) calorimeters at the KEKB and SLAC B-factories
can not stand the planned increase of the luminosity above $20ab^{-1}$.
The CsI(Tl) light yield decreases to about 60\% already at $10ab^{-1}$.
There is also a large increase of PIN diode dark current.
Finally the long decay time of about $1\,\mu sec$ leads to the pile up noise and fake clusters.

The BELLE Collaboration proposes to use pure CsI crystals with a phototetrode readout
and a waveform analysis in the end cap region~\cite{superB}.
The shaping time is reduced from $1\,\mu sec$ to $30\,nsec$. 
The time resolution of better than $1\,nsec$ 
is achieved for energies above $25\,MeV$.
The electronic noise is similar to the present CsI(Tl) calorimeters.
The pure CsI crystals keep more than 90\% of the light output after the irradiation of $7\,kRad$.

   The BaBar Collaboration considers more radiation hard options of LSO or LYSO crystals 
and a liquid Xe calorimeter with the light readout~\cite{babarecal}.
 The LSO and LYSO crystals are 
radiation hard, fast, and dense (see Table~\ref{crystal}). 
They meet perfectly the requirements of the Super
B-factory but their cost is prohibitively high at the moment. 
Liquid Xe is also an attractive option as it is seen in Table~\ref{crystal}.  
The challenge here is the UV light collection. 
BaBar proposes to use WLS fibers and WLS cell coating for an immediate shift
of the light wave length into a region with smaller absorption.  
\begin{table}
\caption{Properties of different scintillators.}
\label{crystal}
\begin{tabular}{|c|c|c|c|} 
 
\hline 
 
\raisebox{0pt}[12pt][6pt]{Scintillator} & 
 
\raisebox{0pt}[12pt][6pt]{CsI(Tl)} & 
 
\raisebox{0pt}[12pt][6pt]{LSO} &
 
\raisebox{0pt}[12pt][6pt]{LiXe} \\
 
\hline
 
\raisebox{0pt}[12pt][6pt]{Density (g/cc)} & 
 
\raisebox{0pt}[12pt][6pt]{4.53} & 
 
\raisebox{0pt}[12pt][6pt]{7.40} & 
 
\raisebox{0pt}[12pt][6pt]{2.95} \\
\hline
\raisebox{0pt}[12pt][6pt]{$X_0$ (cm)} & 
 
\raisebox{0pt}[12pt][6pt]{1.85} & 
 
\raisebox{0pt}[12pt][6pt]{1.14} & 
 
\raisebox{0pt}[12pt][6pt]{2.87} \\\hline
\raisebox{0pt}[12pt][6pt]{$R_M$ (cm) } & 
 
\raisebox{0pt}[12pt][6pt]{3.8} & 
 
\raisebox{0pt}[12pt][6pt]{2.3} & 
 
\raisebox{0pt}[12pt][6pt]{5.7} \\\hline
\raisebox{0pt}[12pt][6pt]{$\lambda$ scint.(nm) } & 
 
\raisebox{0pt}[12pt][6pt]{550} & 
 
\raisebox{0pt}[12pt][6pt]{420} & 
 
\raisebox{0pt}[12pt][6pt]{175} \\\hline
\raisebox{0pt}[12pt][6pt]{$\tau$ scint.(ns) } & 
 
\raisebox{0pt}[12pt][6pt]{680} & 
 
\raisebox{0pt}[12pt][6pt]{47} & 
 
\raisebox{0pt}[12pt][6pt]{4.2,}\\
\raisebox{0pt}[12pt][6pt]{ } & 
 
\raisebox{0pt}[12pt][6pt]{3340} & 
 
\raisebox{0pt}[12pt][6pt]{} & 
 
\raisebox{0pt}[12pt][6pt]{22, 45}\\\hline
\raisebox{0pt}[12pt][6pt]{Photons/MeV } & 
 
\raisebox{0pt}[12pt][6pt]{56k} & 
 
\raisebox{0pt}[12pt][6pt]{27k} & 
 
\raisebox{0pt}[12pt][6pt]{75k}\\\hline
\raisebox{0pt}[12pt][6pt]{Radiation } & 
 
\raisebox{0pt}[12pt][6pt]{} & 
 
\raisebox{0pt}[12pt][6pt]{} & 
 
\raisebox{0pt}[12pt][6pt]{}\\
\raisebox{0pt}[12pt][6pt]{hardness(Mrad)} & 
 
\raisebox{0pt}[12pt][6pt]{0.01} & 
 
\raisebox{0pt}[12pt][6pt]{100} & 
 
\raisebox{0pt}[12pt][6pt]{ -}\\\hline
\raisebox{0pt}[12pt][6pt]{cost(\$/cc)} & 
 
\raisebox{0pt}[12pt][6pt]{3.2} & 
 
\raisebox{0pt}[12pt][6pt]{$\sim 50$} & 
 
\raisebox{0pt}[12pt][6pt]{2.5}\\\hline
\end{tabular}
\end{table}
 
There is a good experience with very large liquid noble gas calorimeters.
For example the $11\,m^3$ LiKr calorimeter at VEPP-4 has an excellent spatial ($\sim 1\,mm$)
and energy ($\sim 3\%/\sqrt{E})$ resolution~\cite{LiKr}.

\subsection{The CMS Lead Tungstate Calorimeter}

The CMS collaboration summarized at this conference their more than 10 year long 
R\&D on the lead tungstate ($PbWO_4$) calorimeter~\cite{cms}.
The choice of $PbWO_4$(Y/Nb) is driven by its small $X_0=0.89\,cm$, small $R_M=2.19\,cm$, 
fast decay time of $\tau\sim10\,nsec$, and a very high radiation hardness above $200\,kGy$.
More than 37.000 crystal have already been produced at the Bogoroditsk (Russia).
In spite of a small light yield of $\sim8\,p.e./MeV$ the excellent energy resolution of 0.51\% 
has been achieved at $120\,GeV$.
Intensive R\&D together with Hamamatsu resulted in excellent APD operated at a gain of 50.
All 120.000 APDs passed a very strict  acceptance test which included a $500\,krad$ irradiation and 
accelerated aging. Vacuum phototriodes (RIA, St.Petersburg) will be used in the endcaps
 because they are more radiation hard. The main challenge for CMS is to finish the 
production of crystals and to maintain the advantages of this approach in the big calorimeter.

\section{Particle Identification}

\subsection{Cherenkov Counters}

\begin{figure}[ht]
\centering
\begin{picture}(550,135)
\put(5,-3){\includegraphics[width=5.5cm]{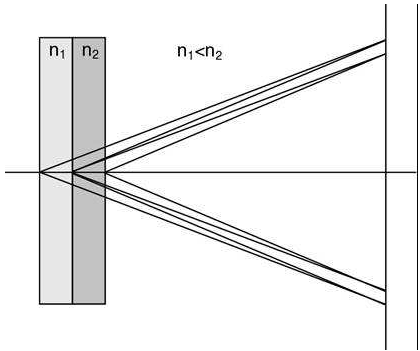}}
\end{picture}
\caption {Principle of the dual radiator Ring Imaging Cherenkov counter . }
\label{rich}
\end{figure}

\begin{figure}[th]
\centering
\begin{picture}(550,180)
\put(0,5){\includegraphics[width=6.5cm]{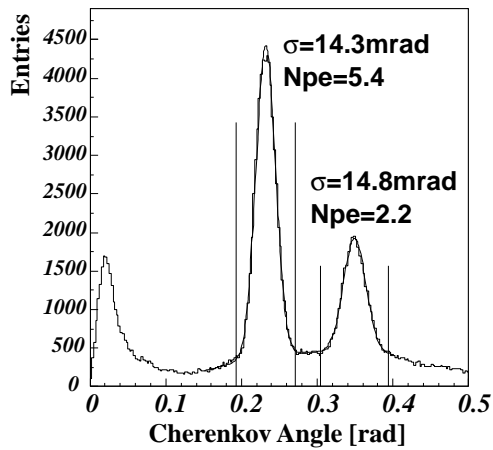}}
\end{picture}
\caption {Distribution of the Cherenkov photon angles from 4$GeV$ pions for a defocusing dual radiator with $n_1=$1.057 and $n_2=$1.027 }
\label{richrings}
\end{figure}

A novel type of proximity focusing RICH counter with a multiple refractive index ($n$) 
aerogel radiator has been developed for the BELLE detector upgrade~\cite{rich}.
The multiple radiator allows to increase the radiator thickness 
and hence the Cherenkov photon yield without degradation in single photon angular resolution.
With the refractive index of the consecutive layers suitably increasing 
in the downstream direction (focusing combination)
one can achieve overlapping of Cherenkov rings from all layers (see Fig.~\ref{rich}).
With the decreasing $n$ (defocusing combination) one can obtain
well separated rings from different layers (see Fig.~\ref{richrings}).


\begin{figure}[ht]
\centering
\begin{picture}(550,310)
\put(0,-3){\includegraphics[width=6.5cm]{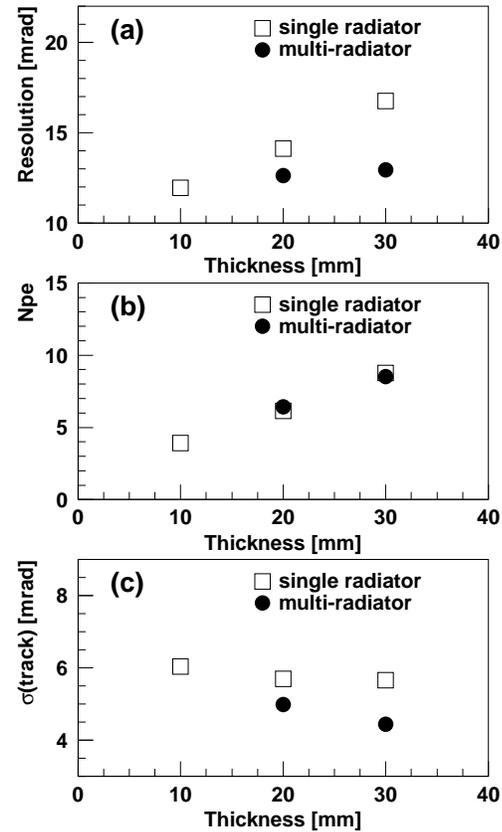}}
\end{picture}
\caption {Single photon resolution (a), number of detected photons (b), and single track Cherenkov angle resolution for single and  multiple focusing radiators for $4GeV$ pions. }
\label{richresult}
\end{figure}

Fig.~\ref{richresult} shows the performance of the detector with the single and multiple
 layer radiators. The number of detected photons is similar in two approaches but
the single photon resolution is much better in the multiple layer configuration.
The Cherenkov angle resolution of $4.5\,mrad$ per track was achieved with the triple
layer radiator. This corresponds to the $5.1\sigma$ $K/\pi$ separation at 4$GeV$.

The radiators with different refraction index layers attached directly at the molecular level 
have been produced at Novosibirsk~\cite{aerogelnov} and in Japan~\cite{rich}.

\begin{figure}[ht]
\centering
\begin{picture}(550,150)
\put(5,-3){\includegraphics[width=6.3cm]{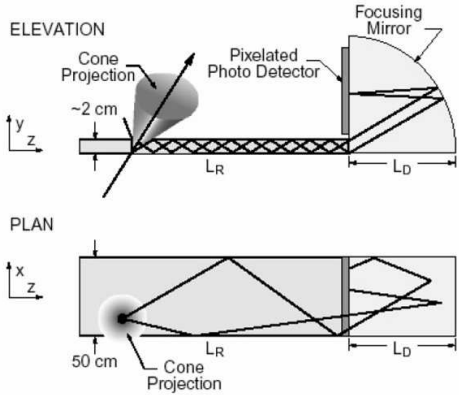}}
\end{picture}
\caption {The principle of FDIRC operation. }
\label{fdirc}
\end{figure}

The BaBar DIRC detector demonstrated an excellent performance. It is natural to 
consider the improved version of this technique for the Super B-factory.
The SLAC and Cincinnati groups develop the Fast Focusing DIRC (FDIRC) detector~\cite{babarecal}.
The idea of this detector is illustrated in Fig.~\ref{fdirc}.
With the accurate time measurement one gets a 3D image of the Cherenkov cone.
In FDIRC the photon detection part is by far smaller than in DIRC.
The development of the pixelated photodetectors with better than $100\,nsec$ time resolution
is a challenging task. The detail studies of Hamamatsu MAPM and 
Burley MCP PM at SLAC give very promising results.  
The FDIRC prototype is ready for tests at SLAC.

In the Time of Propagation (TOP) counter the Cherenkov cone image is reconstructed from the 
coordinate at the quartz bar end and the 
TOP~\cite{top}.
The MCP PM SL10 is developed for TOP together with HPK.  
SL10 has 5 mm pitch and a single photon sensitivity in the 1.5T magnetic field.
The time resolution of $30\,psec$ has been achieved however the cross-talk is still a problem.
The Ga/As photocathodes developed by HPK and Novosibirsk provide enough light for the 
$4\sigma$ $\pi/K$ separation at $4\,GeV$. However the cathode life time is not sufficient yet.
It looses 40\% of quantum efficiency after collecting $350\,mC/cm^2$ 
which corresponds to 6 month operation at the Super B-factory.

\subsection{TOF systems}

\begin{figure}[ht]
\centering
\begin{picture}(550,145)
\put(5,10){\includegraphics[width=6.5cm]{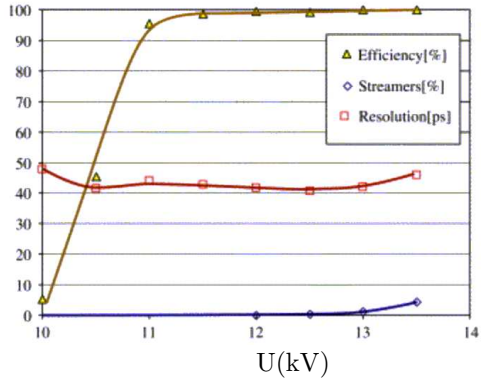}}
\put(100,0){U(kV)}
\end{picture}
\caption {Efficiency (triangles,\%), time resolution  (squares, $nsec$),
 and streamer probability (circles,\%) of MRPC versus applied voltage across 5 gaps (kV). }
\label{alice}
\end{figure}

A multilayer RPC (MRPC) with the excellent time resolution of better than $50\,psec$ 
(see Fig.\ref{alice})
has been developed for the ALICE TOF system~\cite{alicetof}.
It has the  efficiency  of about 99\% at the counting rates as high as few hundred $Hz/cm^2$.
The MRPC has ten $220\,\mu m$ gaps. 
It would be interesting to investigate a possibility to use MRPC for $K_L$ momentum 
measurements in the muon system 
at the Super B-factory.

A time resolution of $48\,psec$ was obtained with a $3\times 3\times 40\,mm^3$ Bicron-418
scintillator read out directly by a $3\times 3\,mm^2$ SiPM without preamplifier~\cite{mephitof}.   
The MIPs were crossing $40\,mm$ in the scintillator. Therefore the signal was 
as big as 2700 pixels in the SiPM with 5625 pixels. The threshold was at 100 pixels.
This approach is very promising for a super high granularity TOF capable to work
in a very high intensity beams for example at FAIR.

\section{Tracking}
 
The Time Projection Chamber (TPC)  is a natural choice for the ILC detector 
central tracker. 
This approach is developed by a large world wide collaboration~\cite{tpc}.
TPC provides continues tracking through a large volume 
with a very small amount of material in front of the ECAL 
($X_0\sim 3\%$ in the barrel region).
The $dE/dx$ resolution of better than $5\%$ helps in particle identification.

\begin{figure}[ht]
\centering
\begin{picture}(550,165)
\put(5,-3){\includegraphics[width=6.5cm]{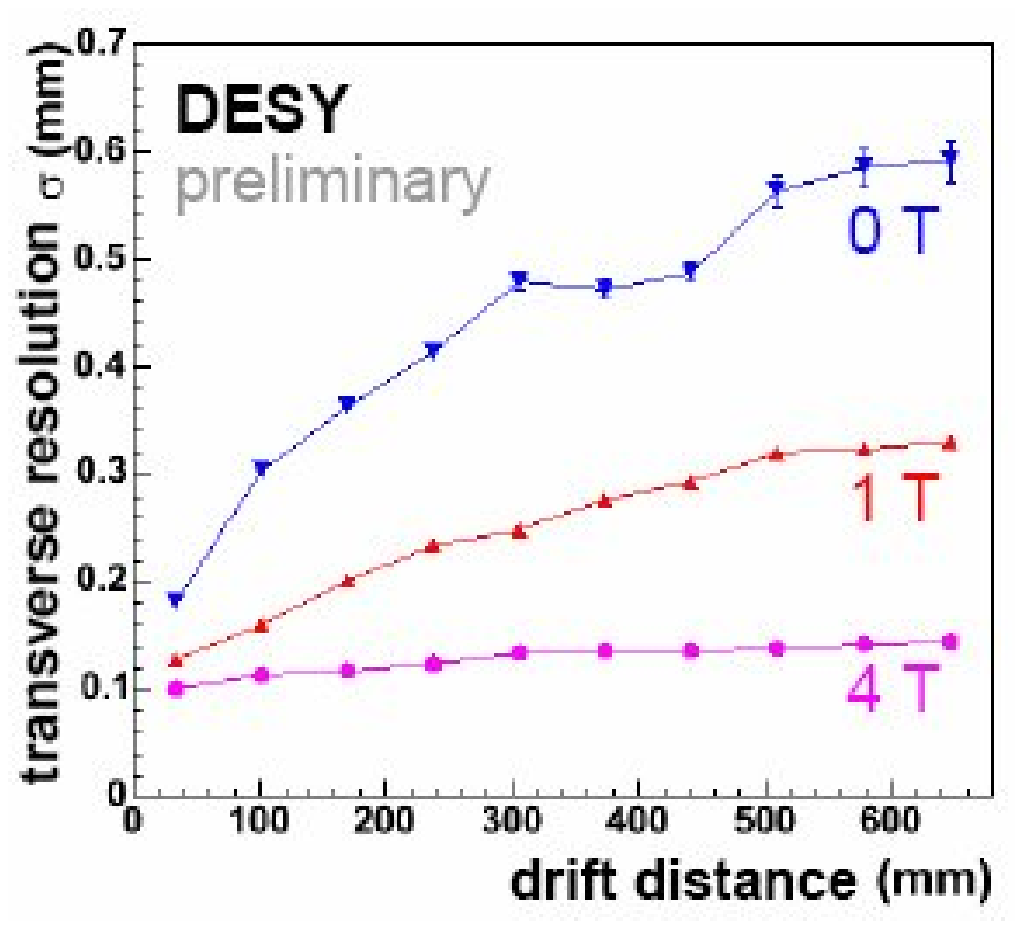}}
\end{picture}
\caption {The transverse resolution dependence on the drift distance for three values of the magnetic field obtained in the TPC with a GEM readout. }
\label{tpc}
\end{figure}

The thrust of the R\&D is in the development of novel micro-pattern
 gas detectors which promise to have a better point 
and two track resolution   than the traditional wire chambers.     
These detectors have smaller ion feedback into the TPC volume.
Micromegas meshes and GEM foils are considered as main candidates.
The spatial resolution of $\sim 100\,\mu m$ was already achieved with GEM after the  $65\,cm$ 
drift in the $4\,T$ field (see Fig.~\ref{tpc}).
Tests at smaller fields demonstrate that a similar resolution can be achieved
 with Micromegas as well. 
The double track resolution of $\sim 2mm$ has been already demonstrated
in small prototypes.
By pitting a resistive foil above the readout pads it is possible 
to spread the signal over several pads. As a result the resolution improves
 up to the diffusion limit.

\begin{figure}[ht]
\centering
\begin{picture}(550,165)
\put(5,-3){\includegraphics[width=6.5cm]{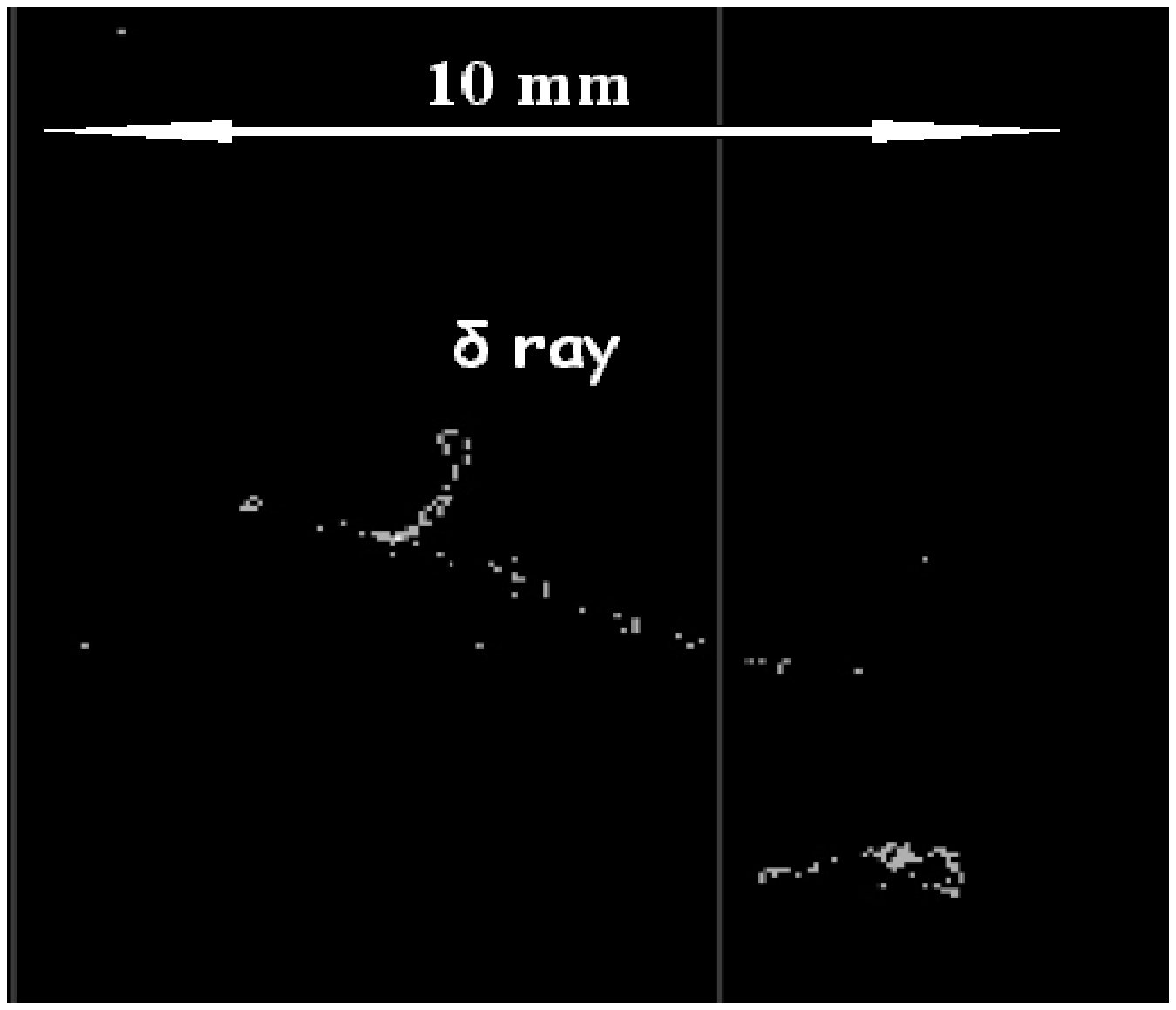}}
\end{picture}
\caption {The transverse resolution dependence on the drift distance for three values of the magnetic field obtained in the TPC with a GEM readout. }
\label{medipix}
\end{figure}

A very exciting approach is a direct TPC readout with the MediPix2 chip~\cite{tpc}.
This CMOS chip contains a square matrix of $256\times 256$ pixels of $55\times 55\,\mu m^2$.
Each pixel is equipped with a low noise preamplifier, discriminator, threshold DAC 
and communication logic.
The extremely high granularity allows to distinguish individual clusters in a track.
Thus the ultimate spatial and dE/dx resolution can be achieved.
Unfortunately the diffusion will severely limit both measurements.
Nice tracks have been recorded by a prototype chamber equipped 
with Micromegas and MediPix2 (see Fig.~\ref{medipix}).
The number of observed clusters ($0.52/mm$ in a He/Isobutane 80/20 mixture) agrees within 15\%
with the expectations.
The next step is to integrate the chip and Micromegas at the postprocessing step
and to add the (drift) time measurement.
Tracks were observed also with a GEM/MediPix2 prototype~\cite{titov}.
A compact all Si tracker is vigorously developed by the US groups~\cite{sitracker}
for the small radius Si Detector for ILC. 
With small detector modules it is possible to reach a very good S/N ratio of about 20,
to have a simple low risk assembly and relatively small amount of material 
of $\sim 0.8\%\,X_0$ per layer including a support structure.
The pattern recognition is a serious issue for the Si tracker 
especially for tracks not coming from the main vertex.

The choice of the central tracker for the Super B-factory depends crucially 
on the expected background which depends on the interaction region design.

In the BELLE study~\cite{superB} the background is expected to increase by a factor
20 from the present values. 
In this case the drift chamber with $13.3\times 16 mm^2$ cells is still adequate for 
the radius above $12.8\,cm$.
Small $5.4\times 5.0 mm^2$ cells are foreseen for the radius between $10.2\,cm$ and $11.6\,cm$.

In the BaBar study~\cite{babartrack} the luminosity term in the background 
extrapolation dominates. Therefore the background estimates are much higher than in the 
BELLE case. A drift chamber can not work in such environment.
Therefore it is proposed to use the all Si tracker up to $R=60\,cm$.
A relatively large amount of material in the Si sensors and support structures
leads to multiple scattering and considerable deterioration of the momentum and mass resolution.
For example the mass resolution in the $B\rightarrow \pi^+\pi^-$ decay mode deteriorates from 
$23MeV$ in case of the drift chamber to $35MeV$ in case of a conservative Si tracker design.
Serious R\&D efforts are required to make the Si tracker thinner.
It should be also demonstrated that the pattern recognition in the Si tracker is good enough.

May be it is possible to develop an alternative solution to the Si tracker.
Using the controlled etching the BINP-CERN group reduced the $Cu$ thickness in GEM foils 
from 5 to $1\mu m$~\cite{lightgem}. This allows to build the light triple GEM chamber with less than 
$0.15\%\,X_0$ including the readout electrode. The light double GEM chamber has even 
smaller thickness. 
The double and triple light GEM chambers were constructed and demonstrated 
identical performance with the standard GEM chambers.
The light GEM chambers have a potential to provide the granularity and spatial resolution
comparable to the Si tracker but with considerably smaller amount of material.
However it is not clear so far how thick a support structure is needed.
A lot of R\&D studies are required to demonstrate a feasibility of this approach.

\section{Conclusions}

The ongoing R\&D should be sufficient to demonstrate the feasibility of detectors for 
the ILC and the Super B-factory.
However there are many promising new ideas which have a potential to improve considerably the 
performance of the detectors and to exploit fully the physics potential of these colliders.
The technologies for practically all detector subsystems are still to be selected
on the basis of the R\&D results.
It is very important to strengthen and to focus the detector R\&D especially 
for the ILC as it was done for the LHC collider.  

\section{Aknowledgments}

This review      
would be impossible without many fruitful discussions with
physicists working on the detector R\&D for the LHC, ILC, and Super B-factory.
In particular we are grateful to  
A.~Bondar, J.~Brau, B.~Dolgoshein, J.~Haba, E.~Popova, F.~Sefkow,
R.~Settles, A.~Smirnitsky.
This work was supported in part by the Russian grants
SS551722.2003.2 and RFBR0402/17307a.

\end{document}